\documentclass[12pt, a4paper]{article}
\usepackage{a4wide,amssymb,amsmath,amscd,}

\newcommand{\Z}{{\mathbb Z}}
\newcommand{\R}{{\mathbb R}}
\newcommand{\C}{{\mathbb C}}

\begin{document}

\topmargin -2pt

\headheight 0pt

\topskip 0mm \addtolength{\baselineskip}{0.20\baselineskip}
\begin{flushright}
{\tt hep-th/0303091} \\
{\tt KIAS-P03016}
\end{flushright}

\vspace{10mm}

\begin{center}
{\large \bf   Theta functions on Noncommutative $T^4$ }\\

\vspace{9mm}

{\sc Hoil Kim}\footnote{hikim@knu.ac.kr}\\

{\it Topology and Geometry Research Center, Kyungpook National University,\\
Taegu 702-701, Korea}\\

\vspace{3mm}

and \\

\vspace{3mm}

{\sc Chang-Yeong Lee}\footnote{cylee@sejong.ac.kr}\\
{\it Department of Physics, Sejong University, Seoul 143-747, Korea}\\

\vspace{18mm}

\end{center}

\begin{center}
{\bf ABSTRACT}
\end{center}
We construct the so-called theta vectors on noncommutative $T^4$,
which correspond to the theta functions on commutative tori with
complex structures. Following the method of Dieng and Schwarz, we
first construct holomorphic connections and then find the
functions satisfying the holomorphic conditions, the theta
vectors. The holomorphic structure in the noncommutative $T^4$
case is given by a $2 \times 2$ complex matrix, and the
consistency requires its off-diagonal elements to be the same. We
also construct the tensor product of these functions
satisfying the consistency requirement.\\

\vfill


\thispagestyle{empty}

\newpage
\section*{I. Introduction}

Classical theta functions have played an important role in the
string loop calculation \cite{jp,gsw}. Recently, noncommutative
geometry \cite{conn} became an important ingredient of string/M
theory (for instance, see \cite{sw99}) starting with the work of
\cite{cds}.

Along this direction, noncommutative torus \cite{cr,rief88} and
its varieties \cite{nct2,nct,hv,t4ours}, and physics on
noncommutative ${\R}^4$ \cite{ns98,gms00,dn01} have been studied
intensively. However, noncommutative tori with complex structures
have been rarely studied \cite{manin,schwarz01,ds02}.
Noncommutative geometry with complex structures has been also
studied recently with algebraic geometry approach for Calabi-Yau
three folds \cite{bl,bs1,bs2}, and for K3 surfaces \cite{kl}.

Classical theta functions can be regarded as state functions over
commutative tori with complex structures. Noncommutative
generalization of this has been initiated in mathematics in the
quantized theta function approach by Manin \cite{manin}, and with
the so-called theta vectors by Schwarz \cite{schwarz01}. In the
physics literature, this has appeared in the context of
noncommutative solitons \cite{mm01,ghs01,hsy02} but mostly in the
so-called integral torus case. Recently, Dieng and Schwarz
\cite{ds02} have computed the theta vectors and their tensor
products on noncommutative $T^2$ explicitly without any
restriction.

In this paper, we follow the method of Dieng and Schwarz and
calculate the theta vectors and their tensor products in the case
of noncommutative $T^4$.
 In section II, we construct modules on
the noncommutative four torus. In section III, we deal with
connections with complex structures. In section IV, we deal with
tensor product of these modules. In section V, we conclude with
discussion.
\\

\section*{II. Modules on noncommutative $T^4$ }\label{module}

In this section, we construct the modules on noncommutative $T^4$
following the method of Rieffel\cite{rief88}.

Recall that $T^d_\theta$ is the deformed algebra of the algebra of
smooth functions on the torus $T^d$ with the deformation parameter
$\theta$, which is a real $d\times d$ anti-symmetric matrix. This
algebra is generated by operators $U_1,\cdots,U_d$ obeying the
following relations
\begin{align*}
U_iU_j=e^{2\pi i \theta_{ij}}U_jU_i \text{ \ and \ }
U_i^*U_i=U_iU_i^*=1, \text{ \ \ } i,j=1,\cdots,d.
\end{align*}
The above relations define the presentation of the involutive
algebra
$${\cal A}_\theta^d=
\{\sum a_{i_1\cdots i_d}U_1^{i_1}\cdots U_d^{i_d}\mid
a=(a_{i_1\cdots i_d})\in {\cal S}({\Z}^d)\}$$ where ${\cal
S}({\Z}^d)$ is the Schwartz space of sequences with rapid decay.

Every projective module over a smooth algebra ${\cal
A}^{d}_{\theta}$ can be represented by a direct sum of modules of
the form ${\cal S}({\R}^p\times{\Z}^q\times F)$, the linear space
of Schwartz functions on ${\R}^p\times{\Z}^q\times F$, where
$2p+q=d$ and $F$ is a finite abelian group.
 The module action is specified
by operators on ${\cal S}({\R}^p\times{\Z}^q\times F)$ and the
commutation relation of these operators should be matched with
that of elements in ${\cal A}^{d}_{\theta}$.

 Recall that there is the dual action of
the torus group $T^d$ on ${\cal A}_\theta^d$ which gives a Lie
group homomorphism of $T^d$ into the group of automorphisms of
${\cal A}_\theta^d$. Its infinitesimal form generates a
homomorphism of Lie algebra $ L$ of $T^d$ into Lie algebra of
derivations of ${\cal A}_\theta^d$. Note that the Lie algebra $L$
is abelian and is isomorphic to ${\R}^d$. Let $\delta:L\rightarrow
{\rm {Der \ }}({\cal A}_\theta^d)$ be the homomorphism. For each
$X\in L$, $\delta(X):=\delta_X$ is a derivation i.e., for $u,v\in
{\cal A}_\theta^d$,
\begin{equation} \label{derv1}
\delta_X(uv)=\delta_X(u)v+u\delta_X(v).
\end{equation}
Derivations corresponding to the generators $\{e_1,\cdots,e_d\}$
of $L$ will be denoted by $\delta_1,\cdots,\delta_d$. For the
generators $U_i$'s of $T_\theta^d$, it has the following property
\begin{equation} \label{derv2}
\delta_i(U_j)=2\pi i\delta_{ij} U_j.
\end{equation}
If $E$ is a projective ${\cal A}_\theta^d$-module, a connection
$\nabla$ on $E$ is a linear map from $E$ to $E\otimes L^*$ such
that for all $X\in L$,
\begin{align} \label{conn1}
\nabla_X(\xi u)=(\nabla_X\xi)u+\xi\delta_X(u),{\rm { \ \ \
}}\xi\in {E}, u\in {\cal A}_\theta^d.
\end{align}
It is easy to see that
\begin{align} \label{conn2}
[\nabla_i,U_j]=2\pi i\delta_{ij} U_j.
\end{align}

We now consider the endomorphisms algebra of a module over ${\cal
A}_\theta^d$. Let $\Lambda$ be a lattice in $G=M\times
\widehat{M}$, where $M={\R}^p\times{\Z}^q\times F$ and
$\widehat{M}$ is its dual. Let $\Phi$ be an embedding map such
that $\Lambda$ is the image of ${\Z}^d$ under the map $\Phi$. This
determines a projective module which will be denoted by
${E}_\Lambda$ \cite{rief88}. The dual lattice of $\Lambda$ can be
defined as
\begin{equation}\label{dual}
\Lambda^\perp:=\{(n,\hat t)\in M\times \widehat{M}\mid
\theta((m,\hat s),(n,\hat t))=<m, \hat t>-<n, \hat s> \in{\Z},
\text{for all }(m,\hat s)\in \Lambda\},
\end{equation}
since in the Heisenberg representation the operators acting on
${E}_\Lambda$ are defined by
\begin{equation}\label{pirep}
{\cal U}_{(m,\hat s)}f(r)=e^{2\pi i <r, \hat s>}f(r+m)
\end{equation}
for $f \in {E}_\Lambda,  \ r \in M .$ Namely, the operators
defined in the dual lattice, ${\cal U}_{(n,\hat t)}$ for $(n,\hat
t)\in\Lambda^\perp$, commute with all the operators defined in the
original lattice, ${\cal U}_{(m,\hat s)}$ for $(m,\hat
s)\in\Lambda$.

It is known that the algebra of endomorphisms on ${E}_\Lambda$,
denoted by $\text{End}_{{\cal A}_\theta}({E}_\Lambda)$, is a
$C^*$-algebra obtained by $C^*$-completion of the space spanned by
operators ${\cal U}_{(n,\hat t)}$, $(n,\hat t)\in\Lambda^\perp$.
The algebra $\text{End}_{{\cal A}_\theta}({E}_\Lambda)$ can be
identified with a noncommutative torus ${\cal
A}_{\tilde{\theta}}$, i.e., ${\cal A}_{\tilde{\theta}}$ is Morita
equivalent to ${\cal A}_\theta$\cite{rief88}.
Recall that a $C^*$-algebra $A$ is said to be Morita equivalent to
$A'$ if $A' \cong \text{End}_A(E)$ for some finite projective
module $E$. In general, a noncommutative torus ${\cal
A}_{\tilde{\theta}}$ is Morita equivalent to ${\cal A}_\theta$ if
$\theta$ and $\tilde{\theta}$ are related by
$\tilde{\theta}=(A\theta+B)(C\theta+D)^{-1}$, where $\left(
\begin{array}{cc} A&B\\C&D\end{array} \right) \in \text{
SO}(d,d|{\Z})$  \cite{rs98}.

In this paper, we consider a projective module of the form ${\cal
S}({\R}^p\times{\Z}^q)\otimes {\cal S}(F)$ with $p=2, \ q=0$.

For the real part, we choose our
embedding map as
\begin{align}\label{phii}
\Phi_{\rm inf}=\begin{pmatrix}\theta_1 + \frac{n_1}{m_1}&0 &0&0\\
                0&0&\theta_2 + \frac{n_2}{m_2}&0\\
                0&1&0&0\\
                0&0&0&1\end{pmatrix} \equiv (x_{ij}),
\end{align}
then using the previous expression for the Heisenberg
representation with $s_1,s_2 \in \R$
\begin{align}
(V_if)(s_1,s_2)=(V_{e_i}f)(s_1,s_2):=\exp(2\pi
i(s_1x_{3i}+s_2x_{4i}))f(s_1+x_{1i},s_2+x_{2i}), \notag
\end{align}
we get
\begin{align}
(V_1f)(s_1,s_2)&=f(s_1 + \theta_1 + \frac{n_1}{m_1}, s_2), \nonumber\\
(V_2f)(s_1,s_2)&=\exp( 2\pi i s_1 )f(s_1, s_2), \nonumber\\
(V_3f)(s_1,s_2)&=f(s_1,s_2 + \theta_2 + \frac{n_2}{m_2} ), \notag\\
(V_4f)(s_1,s_2)&=\exp(2 \pi i s_2)f(s_1,s_2).\nonumber
\end{align}

For the finite part, let $F={\Z}_{m_1}\times{\Z}_{m_2}$, where
${\Z}_{m_i}={\Z}/m_i{\Z}$, ($i=1,2$) and consider the space
${\C}^{m_1}\otimes{\C}^{m_2}$ as the space of functions on
$C({\Z}_{m_1}\times{\Z}_{m_2})$. For all $m_i\in {\Z}$ and $n_i\in
{\Z}/m_i{\Z}$ such that $m_i$ and $n_i$ are relatively prime.
We define the operators $W_i$ on $C({\Z}_{m_1}\times{\Z}_{m_2})$
corresponding to our embedding map
\begin{align}\label{phif}
\Phi_{\rm fin}=\begin{pmatrix} -1&0&0&0\\
                0&0&-1&0\\
                0& \frac{n_1}{m_1}&0&0\\
                0&0&0& \frac{n_2}{m_2}\end{pmatrix}
\end{align}
with $ k_i \in \Z_{m_i} \ (i=1,2) $ as follows
\begin{align}
(W_1f)(k_1,k_2)&=f(k_1-1,k_2), \nonumber\\
(W_2f)(k_1,k_2)&=\exp(2\pi i \frac{n_1 k_1}{m_1})f(k_1,k_2), \nonumber\\
(W_3f)(k_1,k_2)&=f(k_1,k_2-1), \notag\\
(W_4f)(k_1,k_2)&=\exp(2 \pi i
\frac{n_2k_2}{m_2})f(k_1,k_2).\nonumber
\end{align}

Thus, we define operators $U_i=V_i\otimes W_i$ acting on the space
$E_{T}:={\mathcal S}({\mathbb R}^2)\otimes {\mathbb
C}^{m_1}\otimes{\C}^{m_2}$ as
\begin{align} \label{uopr}
(U_1f)(s_1,s_2,k_1,k_2)&=
                 f(s_1  + \theta_1 + \frac{n_1}{m_1},s_2,k_1-1,k_2), \notag \\
(U_2f)(s_1,s_2,k_1,k_2)&=e^{2\pi i(s_1 + \frac{n_1 k_1}{m_1} )}
                 f(s_1,s_2,k_1,k_2), \notag \\
(U_3f)(s_1,s_2,k_1,k_2)&=
                 f(s_1,s_2 + \theta_2 + \frac{n_2}{m_2},k_1,k_2-1), \notag \\
(U_4f)(s_1,s_2,k_1,k_2)&=e^{2\pi i(s_2 + \frac{n_2 k_2}{m_2} )}
                 f(s_1,s_2,k_1,k_2) .
\end{align}
One can now see that they satisfy
\begin{align} \label{Ucomm}
U_2U_1 &=e^{2\pi i \theta_1}U_1U_2, \notag \\
U_4U_3 &=e^{2\pi i \theta_2}U_3U_4,
\end{align}
and otherwise $U_i U_j = U_j U_i $.


In order to find operators which commute with the $U_i$'s, we
recall the definition of the dual lattice $ \Lambda^\perp $:
\[ <m, \hat t>-<n, \hat s> \in{\Z}, \ \
 \text{for all } \ (m,\hat s)\in \Lambda \ \ \text{and} \ \
 (n,\hat t)\in  \Lambda^\perp . \]
If we express the embedding map $\Phi$ as
\begin{align}\label{Tmap}
\Phi=\begin{pmatrix} m& \cdots \\
                \hat s & \cdots \end{pmatrix},
\end{align}
and the embedding map $\Psi$ for the dual lattice as
\begin{align}\label{Smap}
 \Psi=\begin{pmatrix} n& \cdots \\
                \hat t & \cdots \end{pmatrix} ,
\end{align}
then the duality condition above can be written as
\[ <m,\hat t> - < n,\hat s> = \Phi^t J \Psi \in \Z \]
where
\begin{align}\label{Jdef}
J=\begin{pmatrix} 0&0&1&0\\
                0&0&0&1\\
                -1&0&0&0\\
                0&-1&0&0\end{pmatrix}.
\end{align}
Hence, we obtain the relation between the two embedding maps
\begin{equation} \label{maprel}
\Psi = -J \Phi^{-t} \Z .
\end{equation}
%
%
 Using the above relation, the dual map for the real part is now given by
\begin{align}\label{rpsi}
\Psi_{\rm inf}=\begin{pmatrix}\ 0 &  \frac{1}{m_1} &0&0\\
                0&0&0& \frac{1}{m_2}\\
                \frac{1}{m_1 \theta_1 + n_1}&0&0&0\\
                0&0&  \frac{1}{m_2 \theta_2 + n_2}&0\end{pmatrix},
\end{align}
and the finite part is given by
\begin{align}\label{fpsi}
\Psi_{\rm fin}=\begin{pmatrix} 0&-a_1&0&0\\
                0&0&0&-a_2\\
                \frac{1}{m_1}&0&0&0\\
                0&0&  \frac{1}{m_2}&0 \end{pmatrix} .
\end{align}
Here, $a_i \in \Z$ and $a_i n_i -b_i m_i  = 1$ for some $b_i \in
\Z \ (i=1,2)$.

The generators of operators corresponding to the embedding map for
the dual lattice are thus defined by
\begin{align} \label{zop}
(Z_1f)(s_1,s_2,k_1,k_2)&=e^{2\pi i(\frac{s_1}{m_1 \theta_1 +n_1}
  + \frac{k_1}{m_1} )} f(s_1,s_2,k_1,k_2), \notag \\
(Z_2f)(s_1,s_2,k_1,k_2)&= f(s_1 + \frac{1}{m_1},s_2,k_1 -a_1,k_2), \notag \\
(Z_3f)(s_1,s_2,k_1,k_2)&=e^{2\pi i(\frac{s_2}{m_2 \theta_2 +n_2}
  + \frac{k_2}{m_2} )}  f(s_1,s_2,k_1,k_2), \notag \\
(Z_4f)(s_1,s_2,k_1,k_2)&=  f(s_1,s_2 + \frac{1}{m_2},k_1,k_2-a_2).
\end{align}
Here,
\begin{align} \label{zcomm}
Z_1Z_2 &=e^{2\pi i \theta'_1}Z_2Z_1, \notag \\
Z_3Z_4 &=e^{2\pi i \theta'_2}Z_4Z_3,
\end{align}
where \begin{equation} \label{thetap}
\theta'_i = \frac{a_i
\theta_i + b_i}{m_i \theta_i + n_i } , \ \ i=1,2 ,
\end{equation}
 and otherwise $Z_i Z_j = Z_j Z_i$. One can check that the
$Z_i$'s commute with the $U_i$'s, i.e., $ U_iZ_j =Z_j U_i .$


\section*{III. Connections with complex structures }

In the previous section, connections on a projective ${\cal
A}_\theta^d$-module satisfies the condition (\ref{conn2})
\begin{align*}
[\nabla_i,U_j]=2\pi i\delta_{ij} U_j .
\end{align*}
A connection $ \nabla_i$ is called a constant curvature connection
if $ [ \nabla_i , \nabla_j ] = i F_{ij}$ for  constants $F_{ij}$.
 This condition is satisfied if $ \nabla_i $
is expressed as $ \nabla_i = \partial_i - \frac{i}{2}F_{ij} s_j $
where $\partial_i $ is a derivative with respect to $s_i$. Note
that the condition (\ref{conn2}) can be regarded as a
compactification condition. This can be seen by considering an
operator $\overline{X}_i = - \nabla_i$ with which the condition is
expressed as
\begin{align} \label{Xi}
 U_j \overline{X}_i U_j^{-1} &= \overline{X}_i
+ 2 \pi i \delta_{ij}, \end{align}
and this relation is comparable
to a compactification with radius $R_i$,
 $ U_j X_i U_j^{-1} = X_i + 2 \pi  \delta_{ij} R_i .$

We thus let
\begin{align*}
(\overline{X}_i f)(s_1,s_2,k_1,k_2)&=2\pi
iA_{i1}s_1f(s_1,s_2,k_1,k_2)+2\pi
iA_{i2}s_2f(s_1,s_2,k_1,k_2)\\
&-A_{i3}\frac{\partial f(s_1,s_2,k_1,k_2)}{\partial s_1}-
A_{i4}\frac{\partial f(s_1,s_2,k_1,k_2)}{\partial s_2},
\end{align*}
where $A_{ik} \in \R$ are constants. If we denote the embedding
maps as $\Phi_{\text{inf}} \equiv (x_{ij})$ and $\Phi_{\text{fin}}
\equiv (y_{ij}),$ then $U_i$ action is expressed  as
\begin{equation*}
(U_i f)(s_1,s_2,k_1,k_2) = e^{2 \pi i ( s_1 x_{3i}
+ s_2 x_{4i} + k_1 y_{3i} + k_2 y_{4i} )} f ( s_1 +x_{1i},s_2
+x_{2i},k_1 +y_{1i},k_2 +y_{2i}) .
\end{equation*}
The condition (\ref{Xi}) is satisfied if
\begin{equation} \label{embcon}
x_{1i}x_{3i}+ x_{2i}x_{4i} + y_{1i}y_{3i} +y_{2i}y_{4i} =0 ,
\end{equation}
 and
\begin{equation} \label{Aik}
A_{ik} =(\Phi_{\text{inf}}^{-1})_{ik} .
\end{equation}
 The embedding maps  (\ref{phii}), (\ref{phif}) satisfy
the condition (\ref{embcon}), and the condition (\ref{Aik}) gives
\[ ( A_{ik}) =\begin{pmatrix} \frac{1}{\theta_1 + \frac{n_1}{m_1}}&0&0&0\\
                0&0&1&0\\
               0&  \frac{1}{\theta_2 + \frac{n_2}{m_2}}&0&0\\
                0&0& 0&1 \end{pmatrix} . \]
Therefore the following operators specify a constant curvature
connection of right $ T_\theta^4$-module $E_{N,M}$:
\begin{align} \label{connection}
\nabla_1 & = - \frac{2 \pi i s_1 }{\theta_1 + \frac{n_1}{m_1}} ,
\notag \\
\nabla_2 & =  \frac{\partial}{\partial s_1} , \notag \\
\nabla_3 & = - \frac{2 \pi i s_2 }{\theta_2 + \frac{n_2}{m_2}} ,
\notag \\
\nabla_4 & =  \frac{\partial}{\partial s_2} .
\end{align}
In general, a constant curvature connection can be obtained by
adding some constants: \( \nabla_i \rightarrow \nabla_i + d_i , \
i=1,\cdots,4, \) where $ d_i  \in \R $ are constants.

A complex structure on the module $E_{N,M}$ can be introduced by
fixing $ \overline{\partial}$-connection
\begin{align} \label{holconn}
\overline{\nabla}_1 & = \lambda_{11} \nabla_1 + \lambda_{12}
\nabla_2 + \lambda_{13} \nabla_3 + \lambda_{14} \nabla_4 , \notag \\
 \overline{\nabla}_2 & = \lambda_{21} \nabla_1 + \lambda_{22}
\nabla_2 + \lambda_{23} \nabla_3 + \lambda_{24} \nabla_4 , \notag
\end{align}
where $\lambda_{ij} \in \C .$ \ Choosing an appropriate basis such
that $(\lambda_{ij})$ becomes
\begin{equation*}
\begin{pmatrix} \tau_{11} & 1 & \tau_{12} & 0 \\
          \tau_{21} & 0 & \tau_{22} & 1 \end{pmatrix}
                   =  \begin{pmatrix} \lambda_{12} & \lambda_{14} \\
            \lambda_{22} & \lambda_{24} \end{pmatrix}^{-1}
            \begin{pmatrix}
           \lambda_{11}& \lambda_{12} &
             \lambda_{13} & \lambda_{14} \\
           \lambda_{21} & \lambda_{22} &
           \lambda_{23} & \lambda_{24}
            \end{pmatrix} ,
\end{equation*}
 the $(2 \times 2) $  matrix $\begin{pmatrix} \tau_{11} & \tau_{12}  \\
          \tau_{21}  & \tau_{22}  \end{pmatrix}, \ \tau_{ij} \in \C \ $ represents
          the complex structure of $ T_\theta^4$-module and 1-1 corresponds to
          the complex structure on $ T_\theta^4  $ via $\overline{\partial}$-derivative,
           $ \overline{\delta}_i = \sum_j \lambda_{ij} \delta_j $ where $\delta_j$ is
           defined by (\ref{derv2}) \cite{schwarz01}.

Now we consider holomorphic vectors  in $T_\theta^4$-module. A
vector $\Theta \in E_{N,M} $ is called holomorphic
\cite{schwarz01} if it satisfies
\begin{equation} \label{thetavector}
( \overline{\nabla}_i -c_i ) \Theta = 0 \ \ \text{for} \ \  i=1,2,
\end{equation}
where $ c_i  \in \C $ are constants.
 The above holomorphic
condition for $ f \in E_{N,M} $ now takes the form
\begin{align} \label{holcond}
( \frac{2 \pi i \tau_{11} }{\theta_1 + \frac{n_1}{m_1}} s_1  +
 \frac{2 \pi i \tau_{12} }{\theta_2 + \frac{n_2}{m_2}} s_2
  + c_1 ) f
 & =   \frac{\partial f}{\partial s_1} , \notag \\
 ( \frac{2 \pi i \tau_{21} }{\theta_1 + \frac{n_1}{m_1}} s_1  +
 \frac{2 \pi i \tau_{22} }{\theta_2 + \frac{n_2}{m_2}} s_2
  + c_2 ) f
 & =  \frac{\partial f}{\partial s_2} .
\end{align}
%
In order for the two equations in (\ref{holcond}) to be consistent
$\tau_{ij}$ should satisfy
\begin{equation} \label{holcondt4}
\frac{\tau_{12}}{ \theta_2 + \frac{n_2}{m_2}} = \frac{\tau_{21}}
{\theta_1 + \frac{n_1}{m_1}}  .
\end{equation}
If $ Re \Omega < 0$, Eq.(\ref{holcond}) has $m_1 \times m_2$
linearly independent solutions, the so-called theta vectors
\cite{schwarz01,ds02} on noncommutative $T^4$,
\begin{equation} \label{thetat4}
f_{(\alpha_1 , \alpha_2 )} (s_1 , s_2 , k_1, k_2 ) = \exp [
\frac{1}{2} S^t \Omega S + C^t S ] \delta_{\alpha_1}^{k_1}
\delta_{\alpha_2}^{k_2}
\end{equation}
where $C= \begin{pmatrix} c_1 \\ c_2 \end{pmatrix},
S=\begin{pmatrix} s_1 \\ s_2 \end{pmatrix},  c_i \in \C,  s_i \in
\R,  k_i \in \Z_{m_i} \ (i =1,2), $ and $ \Omega =
\begin{pmatrix} \frac{2 \pi i \tau_{11} }{\theta_1 + \frac{n_1}{m_1}} &
\frac{2 \pi i \tau_{12} }{\theta_2 + \frac{n_2}{m_2}} \\
\frac{2 \pi i \tau_{21} }{\theta_1 + \frac{n_1}{m_1}} & \frac{2\pi
i \tau_{22} }{\theta_2 + \frac{n_2}{m_2}}
\end{pmatrix} $.
\\

%
\section*{IV. Tensor product}

In this section we consider a tensor product of two bimodules.
Tensor product of a $(C,Y)$-bimodule $E$ and a $(Y,D)$-bimodule
$E'$ over $Y$ results in a $(C,D)$-bimodule $F$ for algebras
$C,Y,D$;
\[ {}_CE_Y \otimes {}_Y E'_{D} = {}_CF_D \]
where the tensor product over $Y$ is obtained by identifying $ ey
\otimes e' \sim e \otimes y e'  \ \ \text{for} \ \ y \in Y, \ e
\in E , e' \in E' $. Note that in this identification, $E$ behaves
as a right $Y$-module and $E'$ behaves as a left $Y$-module. Thus,
we will denote $E_{N,M}$ as a right $T_\theta^4$-module and
$E'_{K,L}$ as a left $T_{\theta}^4$-module. Here, we recall that
$T^4_{\tilde{\theta}}$ is Morita equivalent to $T^4_\theta$ if
$\theta$ and $\tilde{\theta}$ are related by
$\tilde{\theta}=(A\theta+B)(M\theta+N)^{-1}$ where $\left(
\begin{array}{cc} A&B\\M&N \end{array} \right) \in \text{
SO}(4,4|{\Z})$, and
  $T^4_{\tilde{\theta}}  \cong \text{End}_{T^4_\theta}(E)$ for
some finite projective module $E$. In this notation, a right
module $E_{N,M}$ is identified with a left module $E'_{A,M}$. Let
us calculate the tensor product $E_{N,M} \otimes_{T^4_\theta}
E'_{K,L} $ which forms a vector space $S( \R \times
\Z_{n_1l_1+m_1k_1} \times \Z_{n_2l_2+m_2k_2} ) \ $, when each
$N,M,K,L$ is reducible into two blocks represented by the values
$N \sim (n_1,n_2), \ M \sim (m_1,m_2), \ K \sim (k_1,k_2), \ L
\sim (l_1,l_2)$ \cite{kkl01}. For $f(s_1,s_2,\mu_1, \mu_2) \in
E_{N,M}$, and $g(t_1,t_2,\nu_1,\nu_2) \in E'_{K,L} $ where
$s_i,t_i \in \R, \ \mu_i \in \Z_{m_i}, \ \nu_i \in \Z_{l_i} \
(i=1,2) $ the actions of $U_i \in T^4_\theta$ and $Z_i \in
T^4_{\tilde{\theta}}$ are given as follows. \\
The right $U_i$ actions on $E_{N,M}$ are defined as
\begin{eqnarray} \label{URactions}
(U_1f)(s_1,s_2,\mu_1,\mu_2)&=&
                 f(s_1  + \theta_1 + \frac{n_1}{m_1},s_2,\mu_1-1,\mu_2), \notag \\
(U_2f)(s_1,s_2,\mu_1,\mu_2)&=& e^{2\pi i(s_1 + \frac{n_1
  \mu_1}{m_1} )}   f(s_1,s_2,\mu_1,\mu_2), \notag \\
(U_3f)(s_1,s_2,\mu_1,\mu_2)&=&
                 f(s_1,s_2 + \theta_2 + \frac{n_2}{m_2},\mu_1,\mu_2-1), \notag \\
(U_4f)(s_1,s_2,\mu_1,\mu_2)&=& e^{2\pi i(s_2 + \frac{n_2
             \mu_2}{m_2} )}  f(s_1,s_2,\mu_1,\mu_2) .
\end{eqnarray}
The left $U_i$ actions on $E'_{K,L}$ are defined as
\begin{eqnarray} \label{ULactions}
(U_1g)(t_1,t_2,\nu_1,\nu_2)&=&
                 g(t_1  - \theta_1 + \frac{k_1}{l_1},t_2,\nu_1-1,\nu_2), \notag \\
(U_2g)(t_1,t_2,\nu_1,\nu_2)&=& e^{2\pi i(t_1 + \frac{k_1
    \nu_1}{l_1} )} g(t_1,t_2,\nu_1,\nu_2), \notag \\
(U_3g)(t_1,t_2,\nu_1,\nu_2)&=&
                 g(t_1,t_2 - \theta_2 + \frac{k_2}{l_2},\nu_1,\nu_2-1), \notag \\
(U_4g)(t_1,t_2,\nu_1,\nu_2)&=& e^{2\pi i(t_2 + \frac{k_2
   \nu_2}{l_2} )} g(t_1,t_2,\nu_1,\nu_2) .
\end{eqnarray}
The left $Z_i$ actions on $E_{N,M}$ are defined as
\begin{eqnarray} \label{ZLactions}
(Z_1f)(s_1,s_2,\mu_1,\mu_2)&=& e^{2\pi i(\frac{s_1}{m_1 \theta_1
                +n_1} + \frac{\mu_1}{m_1} )} f(s_1,s_2,\mu_1,\mu_2), \notag \\
(Z_2f)(s_1,s_2,\mu_1,\mu_2)&=& f(s_1 + \frac{1}{m_1},s_2,\mu_1 -a_1,\mu_2),\notag \\
(Z_3f)(s_1,s_2,\mu_1,\mu_2)&=& e^{2\pi i(\frac{s_2}{m_2 \theta_2 +
            n_2}  + \frac{\mu_2}{m_2} )}  f(s_1,s_2,\mu_1,\mu_2), \notag \\
(Z_4f)(s_1,s_2,\mu_1,\mu_2)&=& f(s_1,s_2 +
         \frac{1}{m_2},\mu_1,\mu_2-a_2),
\end{eqnarray}
where $a_i \in \Z$ and $a_i n_i -b_im_i =1 $ for some $b_i \in \Z
\ (i=1,2) . $

 Following \cite{ds02}, we define the tensor product $f \otimes g \equiv h \in E_{N,M}
\otimes_{T^4_\theta} E'_{K,L} $ as
\begin{align} \label{tensorpd}
h(r_1,r_2,j_1,j_2) =  & \sum_{q_1 \in \Z} \sum_{q_2 \in \Z} f((n_1
+m_1 \theta_1)r_1 + \frac{n_1 +m_1 \theta_1}{m_1}q_1 -
\frac{l_1(n_1 +m_1 \theta_1)}{m_1(n_1l_1 +m_1k_1)}j_1,
\notag \\
& (n_2 +m_2 \theta_2)r_2 + \frac{n_2 +m_2 \theta_2}{m_2}q_2 -
\frac{l_2(n_2 +m_2 \theta_2)}{m_2(n_2l_2 +m_2k_2)}j_2, -q_1 +a_1
j_1, -q_2 +a_2 j_2)
\notag \\
   \cdot
   & g( (n_1 +m_1 \theta_1)r_1 - \frac{k_1 - l_1
\theta_1}{l_1}q_1 + \frac{k_1 -l_1 \theta_1}{n_1l_1 +m_1k_1}j_1,
\notag \\
& (n_2 +m_2 \theta_2)r_2 - \frac{k_2 - l_2 \theta_2}{l_2}q_2 +
\frac{k_2 -l_2 \theta_2}{n_2l_2 +m_2k_2}j_2,
 q_1, q_2)
\end{align}
for $r_i \in \R, \  j_i \in \Z \ (i =1,2) .$ Then, one can check
that
\begin{align*}
& (U_i f) \otimes g  \sim f \otimes (U_i)g , \\
& (Z_i h) \sim  (Z_i f) \otimes g, \\
& h(r_i, j_i + n_il_i + m_i k_i ) = h(r_i, j_i)
\end{align*}
for $i=1,2$. Notice that in the above calculation $Z_i$'s act on
$h$ as  left actions, since $h$ is regarded here as an element of
a left module $E'$. So far, we have only defined the left actions
of $Z_i$ on a right module $E$. Thus, we define left $Z_i$ actions
on a left module $E'$ as
\begin{align*}
(Z_1 g) & =  (U_2 g) , \\
(Z_2 g) & =  (U_1 g) , \\
(Z_3 g) & =  (U_4 g) , \\
(Z_4 g) & =  (U_3 g) ,
\end{align*}
where $ U_i g $ are defined in (\ref{ULactions}).

If  $E_{N,M}$ is a right module expression of $
(T^4_{\tilde{\theta}}, T^4_{\theta} )$-bimodule and $E'_{K,L}$ is
a left module expression of $ (T^4_{\theta}, T^4_{\theta'}
)$-bimodule, then
 one can also show that $h$ belongs to $ E'_{AK + BL,
NL +MK} (\tilde{\theta})$ where $\tilde{\theta} =(A\theta
+B)(M\theta +N)^{-1}$ and $\theta =(K\theta' +D)(L\theta'
+C)^{-1}$ with $CK-DL \sim 1, \ AN-BM \sim 1 , $ when each
$A,B,M,N,K,D,L,C$ is reducible into 2 blocks in the sense that we
described earlier.

Let us consider the tensor product (\ref{tensorpd}) between the
two theta vectors, $f\in E_{N,M}$ and $g \in E'_{K,L}$,
\begin{align} \label{thetat4p}
f_{(\alpha_1 , \alpha_2 )} (s_1 , s_2 , \mu_1, \mu_2 ) &= \exp [
\frac{1}{2} S^t \Omega S + C^t S ] \delta_{\alpha_1}^{\mu_1}
\delta_{\alpha_2}^{\mu_2}, \notag \\
g_{(\beta_1 , \beta_2 )} (t_1 , t_2 , \nu_1, \nu_2 ) &= \exp [
\frac{1}{2} T^t \Omega' T + {C'}^t T ] \delta_{\beta_1}^{\nu_1}
\delta_{\beta_2}^{\nu_2} ,
\end{align}
%
where $ \  C= \begin{pmatrix} c_1 \\ c_2 \end{pmatrix}, \  C'=
\begin{pmatrix} c'_1 \\ c'_2 \end{pmatrix}, \ S=\begin{pmatrix} s_1 \\
s_2 \end{pmatrix}, \ T=\begin{pmatrix} t_1 \\
t_2 \end{pmatrix}, \  c_i, c'_i \in \C, \  s_i,t_i \in \R, \
  \mu_i \in \Z_{m_i}, \\
\hspace*{.5cm}   \nu_i \in \Z_{l_i} \ (i=1,2), \
 \Omega =
\begin{pmatrix} \frac{2 \pi i \tau_{11} }{\theta_1 + \frac{n_1}{m_1}} &
\frac{2 \pi i \tau_{12} }{\theta_2 + \frac{n_2}{m_2}} \\
\frac{2 \pi i \tau_{21} }{\theta_1 + \frac{n_1}{m_1}} & \frac{2\pi
i \tau_{22} }{\theta_2 + \frac{n_2}{m_2}}
\end{pmatrix} , \ \text{and} \
\Omega' =
\begin{pmatrix} \frac{2 \pi i \tau'_{11} }{-\theta_1 + \frac{k_1}{l_1}} &
\frac{2 \pi i \tau'_{12} }{-\theta_2 + \frac{k_2}{l_2}} \\
\frac{2 \pi i \tau'_{21} }{-\theta_1 + \frac{k_1}{l_1}} &
\frac{2\pi i \tau'_{22} }{-\theta_2 + \frac{k_2}{l_2}}
\end{pmatrix}
$.
\\

The resulting function now takes the form
\begin{align*} \label{tensorh1}
h_{\alpha_1, \alpha_2, \beta_1, \beta_2}(r_1,r_2,j_1,j_2) =  &
\sum_{q_1 \in \Z} \sum_{q_2 \in \Z} \exp ( \frac{1}{2}
\widetilde{\cal A}^t \Omega \widetilde{\cal A} + C^t
\widetilde{\cal A} ) \delta_{\alpha_1,\alpha_2}^{-q_1 +a_1j_1,-q_2 +a_2j_2} \\
& \cdot \exp ( \frac{1}{2} {\widetilde{\cal A}}'{}^t \Omega'
\widetilde{\cal A}' + {C'}^t \widetilde{\cal A}' )
\delta_{\beta_1,\beta_2}^{q_1 , q_2 }
\end{align*}
where $r_i \in \R , \ j_i \in \Z_{n_il_i +m_ik_i} \ (i=1,2) $ and
\begin{align*}
\widetilde{\cal A} & = \begin{pmatrix} (n_1 +m_1 \theta_1)r_1 +
\frac{n_1 +m_1 \theta_1}{m_1}q_1 - \frac{l_1(n_1 +m_1
\theta_1)}{m_1(n_1l_1
+m_1k_1)}j_1 \\
 (n_2 +m_2 \theta_2)r_2 + \frac{n_2 +m_2 \theta_2}{m_2}q_2 -
\frac{l_2(n_2 +m_2 \theta_2)}{m_2(n_2l_2 +m_2k_2)}j_2
\end{pmatrix} , \\
\widetilde{\cal A}' & = \begin{pmatrix} (n_1 +m_1 \theta_1)r_1 -
\frac{k_1 - l_1 \theta_1}{l_1}q_1 + \frac{k_1 -l_1
\theta_1}{n_1l_1
+m_1k_1}j_1 \\
 (n_2 +m_2 \theta_2)r_2 - \frac{k_2 -l_2 \theta_2}{l_2}q_2 +
\frac{k_2 -l_2 \theta_2}{n_2l_2 +m_2k_2}j_2
\end{pmatrix}  .
\end{align*}
  From the delta function relations, we rewrite $q_i$ as \( q_i = p_i + u_i
\frac{m_i l_i}{v_i} , \ u_i \in \Z \ (i=1,2) \) for some integers
 $p_i$ where $ v_i = \text{g.c.d.}(m_i,l_i) $.
 Then, $h$ can be written as
\begin{align*}
h_{\alpha_1, \alpha_2, \beta_1, \beta_2}(r_1,r_2,j_1,j_2) = &
\sum_{u_1 \in \Z} \sum_{u_2 \in \Z} \exp ( \frac{1}{2}( {\cal A} +
{\cal U} )^t \Omega ({\cal A} + {\cal U})
 + C^t {\cal U} + C^t {\cal A} \\
 &   + \frac{1}{2}( {\cal A}' + {\cal U}' )^t \Omega' ({\cal A}' +
  {\cal U}' ) + {C'}^t {\cal U}' + {C'}^t {\cal A}' ) ,
\end{align*}
where
\begin{align*}
 {\cal A} & = \begin{pmatrix} (n_1 +m_1 \theta_1)r_1 + \frac{n_1
+m_1 \theta_1}{m_1}p_1 - \frac{l_1(n_1 +m_1 \theta_1)}{m_1(n_1l_1
+m_1k_1)}j_1 \\
 (n_2 +m_2 \theta_2)r_2 + \frac{n_2 +m_2 \theta_2}{m_2}p_2 -
\frac{l_2(n_2 +m_2 \theta_2)}{m_2(n_2l_2 +m_2k_2)}j_2
\end{pmatrix} , \\
 {\cal A}' & = \begin{pmatrix} (n_1 +m_1 \theta_1)r_1 - \frac{k_1 - l_1
\theta_1}{l_1}p_1 + \frac{k_1 -l_1 \theta_1}{n_1l_1
+m_1k_1}j_1 \\
 (n_2 +m_2 \theta_2)r_2 - \frac{k_2 -l_2 \theta_2}{l_2}p_2 +
\frac{k_2 -l_2 \theta_2}{n_2l_2 +m_2k_2}j_2
\end{pmatrix}  , \\
 {\cal U} & = \begin{pmatrix} \frac{(n_1 +m_1 \theta_1)l_1}{v_1}u_1 \\
            \frac{(n_2 +m_2 \theta_2)l_2}{v_2}u_2  \end{pmatrix} , \ \
  {\cal U}'  = \begin{pmatrix} -\frac{(k_1-l_1\theta_1)m_1}{v_1}u_1 \\
  -\frac{(k_2-l_2\theta_2)m_2}{v_2}u_2
 \end{pmatrix} .
\end{align*}
This can be decomposed into two parts, including the classical
theta function,
\begin{equation}
h = {\cal \vartheta}({\cal T},{\cal Z}) \xi (r_1,r_2,j_1,j_2).
\end{equation}
Here, the classical theta function $\vartheta$ is given by
 \[ \vartheta ({\cal T},{\cal Z}) = \sum_{u_1, u_2 \in \Z} \exp ( \pi i U^t {\cal T} U
 + 2 \pi i {\cal Z}^t U ), \]
where \begin{align*}
  U & = \begin{pmatrix} u_1 \\ u_2 \end{pmatrix}, \
 {\cal T}  =
\begin{pmatrix} {\cal B}_1 {\cal O}_{11} {\cal B}_1 + {\cal B'}_1 {\cal O'}_{11} {\cal B'}_1
 & {\cal B}_1 {\cal O}_{12} {\cal B}_2 + {\cal B'}_1 {\cal O'}_{12} {\cal B'}_2 \\
 {\cal B}_2 {\cal O}_{21} {\cal B}_1 + {\cal B'}_2 {\cal O'}_{21} {\cal B'}_1
  & {\cal B}_2 {\cal O}_{22} {\cal B}_2 + {\cal B'}_2 {\cal O'}_{22} {\cal B'}_2
\end{pmatrix}   ,  \\
     {\cal Z} & = \begin{pmatrix} {\cal B}_1 {\cal O}_{11} {\cal A}_1
     + {\cal B}_1 {\cal O}_{12} {\cal A}_2
     + {\cal B'}_1 {\cal O'}_{11} {{\cal A}'}_1
      + {\cal B'}_1 {\cal O'}_{12} {{\cal A}'}_2
  \\
   {\cal B}_2 {\cal O}_{21} {\cal A}_1
     + {\cal B}_2 {\cal O}_{22} {\cal A}_2
     + {\cal B'}_2 {\cal O'}_{21} {{\cal A}'}_1
      + {\cal B'}_2 {\cal O'}_{22} {{\cal A}'}_2
  \end{pmatrix}
  + \frac{1}{2 \pi i} \begin{pmatrix}
    c_1 {\cal B}_1 + {c'}_1 {\cal B'}_1 \\
    c_2 {\cal B}_2 + {c'}_2 {\cal B'}_2
  \end{pmatrix}
   \end{align*}
   with
  \(  {\cal O} = \frac{1}{2 \pi i} \Omega , \
     {\cal O'} = \frac{1}{2 \pi i} \Omega' , \
   {\cal B}  = \begin{pmatrix}
    {\cal B}_1  \\   {\cal B}_2   \end{pmatrix} =
\begin{pmatrix} \frac{(n_1 +m_1 \theta_1)l_1}{v_1} \\
            \frac{(n_2 +m_2 \theta_2)l_2}{v_2}  \end{pmatrix} ,  \
     {\cal B}' = \begin{pmatrix}
    {\cal B'}_1 \\   {\cal B'}_2    \end{pmatrix}
      =  \begin{pmatrix} -\frac{(k_1-l_1\theta_1)m_1}{v_1} \\
  -\frac{(k_2-l_2\theta_2)m_2}{v_2}
 \end{pmatrix}    , \)
 and the function $\xi $ is given by
\[ \xi(r_1,r_2,j_1,j_2) = \exp ( \frac{1}{2} {\cal A}^t \Omega {\cal A} + \frac{1}{2}
{{\cal A}'}^t \Omega' {\cal A}' + C^t {\cal A} + {C'}^t {\cal A}')
 . \]

Requiring that $E_{N,M}$ and $ E'_{K,L}$ have the same complex
structure for consistency of the tensor product, i.e.,
$(\tau_{ij}) = (\tau'_{ij})$, and the consistency condition
(\ref{holcondt4}) the resulting function $h$ becomes
\begin{equation}
h_{\alpha_1, \alpha_2, \beta_1, \beta_2} (r_1,r_2,j_1,j_2) =
\sum_{\gamma_1, \gamma_2} c_{\alpha_1, \alpha_2, \beta_1,
\beta_2}^{\gamma_1, \gamma_2} \varphi_{\gamma_1, \gamma_2}
(r_1,r_2,j_1,j_2).
\end{equation}
Here, the function $\varphi_{\gamma_1,\gamma_2} $ is given by
\[
 \varphi_{\gamma_1,\gamma_2} (r_1,r_2,j_1,j_2)  =
  \exp ( \pi i R^t \widetilde{\cal O}R + (C + C')^t \widetilde{R}
) \delta_{\gamma_1}^{j_1} \delta_{\gamma_2}^{j_2} \]
 with $ R=  \begin{pmatrix} r_1 \\ r_2 \end{pmatrix}, \
\widetilde{\cal O} =
\begin{pmatrix} \frac{\tau_{11}(n_1 +m_1 \theta_1)(n_1l_1
+m_1k_1)}{(k_1 -l_1 \theta_1)} & \frac{\tau_{12}(n_1 +m_1
\theta_1)(n_2l_2 +m_2k_2)}{(k_2 -l_2 \theta_2)}
  \\
  \frac{\tau_{21}(n_2 +m_2 \theta_2)(n_1l_1
+m_1k_1)}{(k_1 -l_1 \theta_1)} &
 \frac{\tau_{22}(n_2 +m_2
\theta_2)(n_2l_2 +m_2k_2)}{(k_2 -l_2 \theta_2)}
\end{pmatrix} , $
 $
\widetilde{R} = \begin{pmatrix} (n_1 +m_1 \theta_1)r_1   \\
 (n_2 +m_2 \theta_2)r_2
\end{pmatrix} , $ and the constants
$ c_{\alpha_1, \alpha_2, \beta_1, \beta_2}^{\gamma_1, \gamma_2}$
are given by
\[
 c_{\alpha_1, \alpha_2, \beta_1, \beta_2}^{\gamma_1,
\gamma_2}  = \vartheta (\Xi, \Lambda ) e^{\cal K}  \] where
\[ {\cal K}= \pi i {\widetilde{Q}}^t \widetilde{P} + C^t \widetilde{M}  + {C'}^t \widetilde{L}  , \]
with
\[ \widetilde{Q} = \begin{pmatrix} \tau_{11} p_1 + \tau_{12} p_2
   - \frac{\tau_{11}  l_1 j_1} {n_1l_1 +m_1k_1}
    - \frac{\tau_{12}  l_2 j_2} {n_2l_2 +m_2k_2} \\
\tau_{21} p_1 + \tau_{22} p_2
   - \frac{\tau_{21}  l_1 j_1} {n_1l_1 +m_1k_1}
    - \frac{\tau_{22}  l_2 j_2} {n_2l_2 +m_2k_2}
\end{pmatrix} , \
    \widetilde{P} = \begin{pmatrix} \frac{n_1l_1 +m_1k_1}{l_1m_1} p_1
     -\frac{j_1}{m_1} \\
    \frac{n_2l_2 +m_2k_2}{l_2m_2} p_2
     -\frac{j_2}{m_2}
\end{pmatrix} , \]
\[  \tilde{M} = \begin{pmatrix}  \frac{n_1
+m_1 \theta_1}{m_1}p_1 - \frac{l_1(n_1 +m_1 \theta_1)}{m_1(n_1l_1
+m_1k_1)}j_1 \\
  \frac{n_2 +m_2 \theta_2}{m_2}p_2 -
\frac{l_2(n_2 +m_2 \theta_2)}{m_2(n_2l_2 +m_2k_2)}j_2
\end{pmatrix}   , \
    \tilde{L} = \begin{pmatrix} - \frac{k_1 -
l_1 \theta_1}{l_1}q_1 + \frac{k_1 -l_1 \theta_1}{n_1l_1
+m_1k_1}j_1 \\
 - \frac{k_2 -l_2 \theta_2}{l_2}q_2 +
\frac{k_2 -l_2 \theta_2}{n_2l_2 +m_2k_2}j_2
\end{pmatrix} ,
   \]
and
\[ \vartheta (\Xi,\Lambda) = \sum_{u_1, u_2 \in \Z} \exp ( \pi i U^t \Xi U
 + 2 \pi i {\Lambda}^t U ),  \]
with
\begin{align*}
 U & = \begin{pmatrix} u_1 \\ u_2 \end{pmatrix}  , \ \
        \Xi =   \begin{pmatrix} \frac{\tau_{11}  (n_1l_1
+m_1k_1)l_1 m_1}{v_1^2} & \frac{\tau_{12}  (n_1l_1 +m_1k_1)l_2
m_2}{v_1 v_2} \\
\frac{\tau_{21}  (n_2l_2 +m_2k_2)l_1 m_1}{v_1 v_2} &
\frac{\tau_{22} (n_2l_2 +m_2k_2)l_2 m_2}{v_2^2}
\end{pmatrix} , \\
  \Lambda & = \begin{pmatrix} \frac{ \tau_{11} }{v_1} ( (l_1n_1 +m_1k_1)
 p_1 - l_1 j_1) + \frac{ \tau_{12} }{v_1} ( l_1 n_1 +m_1k_1)
 ( p_2 -\frac{l_2 j_2}{l_2n_2 +m_2k_2} ) \\
 \frac{ \tau_{21} }{v_2}  ( l_2 n_2 +m_2k_2)
 ( p_1   -\frac{l_1 j_1}{l_1n_1 +m_1k_1} )
+ \frac{ \tau_{22} }{v_2} ( (l_2n_2 +m_2k_2)
 p_2 - l_2j_2)
 \end{pmatrix} \\
      &  \ \  +  \frac{1}{2 \pi i} \begin{pmatrix}
    c_1 {\cal B}_1 + {c'}_1 {\cal B'}_1 \\
    c_2 {\cal B}_2 + {c'}_2 {\cal B'}_2
  \end{pmatrix} ,
  \end{align*}
where $c_i, c_i' , {\cal B}_i, {\cal B}_i' $ are the same as given
before.

Notice that the function $\varphi_{\gamma_1,\gamma_2}
(r_1,r_2,j_1,j_2)$ is a theta vector (\ref{thetat4p}) belongs to $
{E'}_{AK+BL,NL+MK}(\tilde{\theta})$ with $\tilde{\theta} = \frac{A
\theta +B}{M\theta +N}$ as we expected.
\\



\section*{V. Discussion}

In this paper, we first construct a module on noncommutative $T^4$
and its dual. Then we define the complex structure on this module
and construct the theta vector which is a solution for a
holomorphic connection. We then consider the tensor product of the
theta vectors satisfying the consistency requirement.

Here, we want to notice what has not been apparent in the
noncommutative $T^2$ case \cite{ds02}. When we require the
holomorphic condition (\ref{thetavector}), the symmetry appears
not in the complex structure itself but in the $\Omega$-matrix
which appears in the theta vector (\ref{thetat4}), i.e.,
$\Omega_{12}=\Omega_{21}$ instead of $\tau_{12} = \tau_{21} $ in
the commutative 4-torus case. We consider that this difference
comes from noncommutativity.

As in the noncommutative $T^2$ case, the tensor products of
modules on the noncommutative 4-torus with complex structures
become very restrictive in order to satisfy the consistency
requirement. We consider this consistency requirement as another
aspect of noncommutativity compared with the commutative case in
which there is no restriction.

So far, theta functions on noncommutative tori have not been
utilized in the physics literature except for the integral torus
case \cite{mm01,ghs01,hsy02}. With the theta functions on
noncommutative tori without any restriction, one can hope to
explore the physical states on noncommutative tori in more general
cases.
\\


\vspace{5mm}

\noindent
{\Large \bf Acknowledgments}

\vspace{5mm}
\noindent
Most part of the work was done during
authors' visit to KIAS. The authors would like to thank KIAS for
its kind hospitality. This work was supported by KOSEF
Interdisciplinary Research Grant No. R01-2000-000-00022-0.



\end{document}